\documentclass[aps,prd,preprint,superscriptaddress,showpacs,showkeys]{revtex4}
\usepackage{amssymb,amsmath}
\usepackage{graphicx} 
\usepackage{dcolumn}  
\usepackage{epsfig}   
\usepackage{pstricks}
\usepackage{ifpdf}
\newcommand{\Tr}{ {\mathrm{Tr}\, }}

\begin{document}


\title{On the Casimir operator dependences of QCD amplitudes}


%
%

\author{T. Grandou}
\affiliation{Universit\'{e} de Nice-Sophia Antipolis,\\ Institut Non Lin\'{e}aire de Nice, UMR CNRS 7335; 1361 routes des Lucioles, 06560 Valbonne, France.}
\email[]{Thierry.Grandou@inln.cnrs.fr}


\date{\today}

\begin{abstract} In eikonal and quenched approximations at least, it is argued that the strong coupling fermionic QCD amplitudes based on the newly discovered {\textit{effective locality}} property, depart from a dependence on the sole $SU_c(3)$ quadratic Casimir operator, evaluated over the fundamental gauge group representation: A definite dependence on the cubic $SU(3)$ Casimir operator takes place. \par
This result, in contradistinction with Perturbation Theory, but also with a number of non-perturbative approaches such as the MIT Bag, the Stochastic Vacuum Models, and Lattice simulations, accounts, instead, for the full algebraic content of the rank-2 $SU_c(3)$-Lie algebra.\end{abstract}


\maketitle

\section{\label{SEC:1}Introduction}

In some recent articles~\cite{QCD1,QCD-II,QCD5, QCD6}, a new non-perturbative property, which bears on fermionic Green's functions of QCD, has been demonstrated: {\textit{effective locality}}. That property
can be phrased as follows. For any Quark/Quark (or Anti-Quark) scattering amplitude, the
full gauge-invariant sum of cubic and quartic vectorial gluonic interactions,
fermionic loops included, results in a local contact-type interaction, and this local interaction is
induced by a tensorial field structure antisymmetric both in Lorentz and
color indices. This is a non-expected result because, ordinarily, integrations of elementary degrees of freedom result in highly non-local and non-trivial structures;  the
`effective locality' denomination, which sounds like an {\textit{oxymoron}}, accounts for this
rather unusual circumstance.

If effective locality is a sound, relevant property of QCD, then its consequences, even examined `at tree level', should exhibit admissible as well as new aspects of the confined phase of QCD; and so far, it seems to be so~\cite{QCD-II, QCD5, QCD6}.

In Ref.~\cite{QCD6}, a general form of the QCD fermionic amplitudes is displayed as a finite sum of finite products of {\textit{Meijer's special functions}}, in agreement with most general and theoretical expectations~\cite{Ferrante2011}. Remarkably enough, within one and the same expression, these amplitudes are able to show up an explicit link between a {\textit{partonic content}} and a hadronic non-perturbative component in accord, this time, with the $AdS_5/QCD$ light-cone approach of Ref.\cite{deTeramond2012}.

However, the analysis presented in Ref.\cite{QCD6} is carried out at eikonal and quenched approximations. Soon it will become important to relax these approximations, not only for the sake of preserving unitarity, but also in order to explore larger distances: Effective locality, in effect, clearly differentiates QCD from the pure Yang Mills situation. In particular (as noticed also by lattice approaches), inclusion of quark loops reveals to be essential to the description of larger distance non-perturbative physics~\cite{QCD5}.

Fortunately enough, there are things that can be learnt already at the level of a quenched analysis: In the current article, we will take advantage of the results worked out in Ref.\cite{QCD6} to address a peculiar, non-perturbative issue which could account for a novel aspect of the QCD non-perturbative phase, such as disclosed by the property of effective locality.

\section{\label{SEC:2} Casimir operator dependences}

In Perturbation Theory applied to QCD, all of $2\rightarrow 2$ scattering process calculations yield  results proportional to either $C_A=N_c$ and/or $C_F={(N_c^2-1)}/{2N_c}$, that is to the quadratic Casimir operator eigenvalue $C_2({\cal{R}})$ over the adjoint and fundamental representations respectively (exceptions may be found in Ref.~\cite{Bali}, where higher dimensional representation spaces were considered, but again, restricted to $C_2$-dependences). The quadratic Casimir operator is $C_2({\cal{R}})=\sum_{a=1}^{N_c^2-1} T_a^2({\cal{R}})$, where the $T_a({\cal{R}})$ denote the $SU_c(3)$ Lie algebra generators of a given representation ${\cal{R}}$. \par
Not only perturbative calculations, but also non-perturbative models of QCD, such as the MIT bag model~\cite{MIT}, the Stochastic Vacuum Model (SVM)~\cite{Dosch}, and lattice approaches~\cite{Bali,BMuller}, comply with these overall $C_2({\cal{R}})$ dependences, though, sometimes, in quite different ways.

The property of  effective locality shows up for $2n$-point fermionic Green's functions, as an exact, non-approximate property of QCD~\cite{QCD-II}. In the strong coupling regime $g>>1$, evaluating a 2 by 2 Quark/Quark(Anti-Quark)  scattering amplitude with the help of {\textit{Random Matrix Theory}}, one finds a result which (up to renormalization) is proportional to~\cite{QCD6},
\begin{eqnarray}\label{Eq:1}
&& (-16\pi{m^2\over E^2})^N\sum_{\mathrm{monomials}}(\pm 1)\,\Tr\,C\int{\rm{d}}p_1\  ..\ {\rm{d}}p_{N(N-1)/2}\ f(p_1,\dots, p_{N(N-1)/2})\nonumber\\  & & \!\!  \times\  \prod^{\sum q_i=N(N-1)/2}_{1\leq i\leq N}\ [1-i(-1)^{q_i}] \,\! \int_0^{+\infty} {\rm{d}}\alpha_1^i\ {\sin[\alpha_1^i({\cal{OT}})_i]\over \alpha_1^i}\int_0^{+\infty} {\rm{d}}\alpha_2^i\ {\sin[\alpha_2^i({\cal{OT}})_i]\over \alpha_2^i} \nonumber\\  & & \quad \times \, G^{23}_{34}\left( \left.{2iN_c}\left({ \alpha_1^i\alpha_2^i \over g\varphi(b) }\right)^2{{\hat{s}}({\hat{s}}-4m^2)\over m^4} \right|
\begin{array}{cccc}
  \frac{3 - 2 q_i}{4}, & \frac{1}{2}, & 1, &  \\
   1, & 1, &  \frac{1}{2}, &  \frac{1}{2}
\end{array}
\right)\ ,
\end{eqnarray}where eikonal and quenching approximations have been used. QCD is here simplified to the case of a single quark species of mass $m$, and $E(=E_1=E_2)$ is the 2 colliding quarks energy in the center of mass system, $p_1=(E,0,0,p)$, $p_2=(E,0,0,-p)$, ${\hat{s}}=(p_1+p_2)^2$. This amplitude can be generalized to the case of $2n$-point fermionic Green's functions~\cite{QCD6, {tgrh}}. The monomials which are understood in the summation are those of a {\textit{Vandermonde determinant}}, comprising $2^{N(N-1)/2}$ terms, 
\begin{eqnarray}\label{Eq:13}
{\cal{P}}(\xi_1,\dots,\xi_{N})=\prod_{1\leq i<j\leq N} |\xi_i-\xi_j| \ .
\end{eqnarray}Each monomial is characterized (not in a unique way) by a given distribution of $q_i$-powers whose sum satisfies the constraint of an equal global degree of $N(N-1)/2$. The monomials share the same algebraic color structure, which is that of the net amplitude. With  $D$, the number of spacetime dimensions, the number $N=D\times(N_c^2-1)$~\cite{halpern} defines the format $N\!\times\! N$, of the orthogonal matrix ${\cal{O}}={\cal{O}}(\dots,p_l,\dots)\ , 1\leq l\leq N(N-1)/2$, introduced so as to diagonalize a real random traceless $N\!\times\! N$ symmetric matrix of spectrum, the $\xi_i, 1\leq i\leq N$. It is by integrating over the ${\xi_i}s$ (and over 2 subsidiary variables) that analytic {\textit{Meijer}}'s special functions  $G^{23}_{34}$ Eq.(\ref{Eq:1}) come about~\cite{QCD6}, in agreement with the most general expectations of Ref.\cite{Ferrante2011}. In one and the same argument, the $G^{23}_{34}$-Meijer functions mix up partonic variables ($m, p_1, p_2, {\hat{s}}$) with the non-perturbative function $\varphi(b)\sim({\mu/{\sqrt{\hat{s}}}})\,e^{-(\mu b)^{2-\xi}}$, where $b=|{\vec{b}}|$ is the impact parameter of the scattering process, and $\mu$, the mass scale necessarily introduced by effective locality~\cite{QCD6}. Also, writing (\ref{Eq:1}), the absolute values of (\ref{Eq:13}) have been dropped, as it can be shown that they do not affect the point made in the current letter~\cite{tgrh}.


\par\medskip
The matter under consideration is in the 2nd line of (\ref{Eq:1}) where $f(..,p_l,.)$ is the joint probability distribution of the $N(N-1)/2$-parameters that specify the orthogonal matrix ${\cal{O}}$, whereas the constant $C$ normalizes that distribution to unity. The matricial part of (\ref{Eq:1}) appears in the 2 {\textit{sine functions}}, while a trace over internal color degrees may be taken.

 One has the $N$-vector of matrices ${\cal{T}}=(T,T,T,T)=(1,1,1,1)\otimes T$, that is $D=4$ copies of the original full set $T$ of $SU_c(3)$ generators, taken in the fundamental representation: $T=\{t_1,t_2,\dots,t_7,t_8\}$, with $t_a={\lambda_a/2}$, the standard Gell-Mann matrices~\cite{Yndurain}.


\par\medskip
As advertised in Ref.\cite{QCD6}, the additional, but unavoidable complexity coming from the orthogonal matrix average  is essential to prevent a trivial result from occuring, and to begin with, it is interesting to observe how such a trivial result would effectively come out. 

Ignoring the $p_l$-dependent orthogonal matrix ${\cal{O}}$ in the second line of (\ref{Eq:1}), one gets for a generic monomial of the sum (\ref{Eq:1}),
\begin{equation}
\pm\Tr\! \prod^{\sum q_i=N(N-1)/2}_{1\leq i\leq N} [1-i(-1)^{q_i}] \int_0^{\infty} {\rm{d}}\alpha_1^i\ {\sin(\alpha_1^i{\cal{T}}_i)\over \alpha_1^i}\int_0^{\infty} {\rm{d}}\alpha_2^i\ {\sin(\alpha_2^i{\cal{T}}_i)\over \alpha_2^i} \, G^{23}_{34}\left(C^{st}(\alpha^i_1\alpha^i_2)^2|\dots\right)\,,
\end{equation}where in $G^{23}_{34}$, the $C^{st}$ can be read off (\ref{Eq:1}). That is,
\begin{equation}\label{4}
\pm\Tr \prod_{1\leq i\leq N} [1-i(-1)^{q_i}] \int_0^{\infty} {\rm{d}}\alpha_1^i\ {\sin\alpha_1^i\over \alpha_1^i}\int_0^{\infty} {\rm{d}}\alpha_2^i\ {\sin\alpha_2^i\over \alpha_2^i} \, G^{23}_{34}\left( \left.C^{st}(\alpha^i_1\alpha^i_2)^2\,\right|
\begin{array}{c}
 a_r(i)\\
 b_s
\end{array}\right) \frac{{{\lambda}}_{{\widehat{i}}}^2}{4}\,.
\end{equation}Expanding the sine-functions, the property ${\cal{T}}_j=t_{{\widehat{j}}}={{\lambda}}_{{\widehat{i}}}/2$ has been used, where ${\widehat{i}}=i\ modulo\ N_c^2-1$.
Now, the following equalities hold,
\begin{equation}\label{nul}
\ \,\prod_{i=1}^N\, {\cal{T}}_j^2=\ \left(\prod_{a=1}^{N_c^2-1=8}\, \frac{1}{4}\lambda_a^2\right)^4=(\frac{1}{4})^{21}\left(\prod_{a=1}^{7}\, \frac{1}{4}\lambda_a^2\right)(\frac{1}{4}\lambda_8^2)^4={\mathbf{0}}_{\,3\times 3}\,(\frac{1}{4}\lambda_8^2)^4=0\,.
\end{equation}The first equalities hold by definition of the $N$-vector of matrices ${\cal{T}}$, and because of the relations $ [\lambda^2_a, \lambda^2_b]=0\, ,\forall a, b=1,2,..,N_c^2-1$; the last equalities can be checked by a direct evaluation at $N_c=3$, using standard Gell-Mann matrices.
\par\medskip
Restoring the relevant dependences on ${\cal{O}}$ now, one can write for a given monomial of the sum (\ref{Eq:1}) the result \cite{tgrh},
\begin{eqnarray}\label{6}
&& \nonumber\pm  (-{16\pi^2 m^2\over E^2})^N\ <\,\prod_{i=1}^N\, [1-i(-1)^{q_i}] \\ &&\times\, \left({{\sqrt{32iN_c}}\,Ep\over {m^2}}\right) \frac{[({\cal{OT}})_i]^{-2}}{g\varphi(b)}\,G^{30}_{03}\!\left( [{  g\varphi(b)\over {\sqrt{512iN_c}}  }{m^2\over Ep}]^2[({\cal{OT}})_i]^{4}\, \biggr|\frac{1}{2}, \frac{3+2q_i}{4},1\!\right)\ >
\end{eqnarray}where the brackets are here to mean an $O_N(\mathbb{R})$-averaged quantity. Equation (\ref{6}) is obtained out of (\ref{Eq:1}) by integration over the $\alpha_J^i,\ J=1,2$ : In the Meijer special functions $G^{23}_{34}$ of  (\ref{Eq:1}), in effect, the sequences of numbers $\{ a_r(i) \}$ and $\{   b_s \}$ (see Eq.(\ref{4})), are such that {\textit{formula}} ${{20.5.}}(7)$ of Ref.~\cite{Erdelyi1954} can be used twice; {\textit{formulae}} 5.3.1(8) and 5.3.1(9) of Ref.\cite{Erdelyi1953} have also been used to reduce an initial Meijer function of $G^{52}_{47}$ to the much simpler one of (\ref{6}), $G^{30}_{03}$.
\par\medskip\noindent
The matrix-valued argument, $z$, of $G^{30}_{03}$ is proportional to a number,
\begin{equation}\label{D4}
\lambda\equiv | (\,{  g\varphi(b)\over {\sqrt{32iN_c}}  }{m^2\over{\sqrt{{\widehat{s}}({\widehat{s}}-4m^2)}}}\,)^{2}|\,, \ \ \ \ \ \ \ \ \varphi(b)\simeq \frac{\mu}{{\sqrt{{\hat{s}}}}}\ e^{-(\mu b)^{2-\xi}}\,,
\end{equation}
which, even at large enough coupling, $g= 10$,  can be very small in the range of moderate sub-energies $\hat{s}$. In that single quark model, taking $m\simeq 5$MeV, and a sub-energy $\hat{s}\simeq 100$ MeV, the simplest non-relativistic  estimations of \cite{QCD-II}, with $\xi={\sqrt{2}}/16$ and $\mu/m=2^{3/4}{\sqrt{2/\xi}}$, leads to an estimation of $\lambda$ on the order of $0.004$, while the matrix elements of $[({\cal{OT}})_i]^{4}$ are themselves less or equal to 1: For all $i,j$, in effect, one has $|{\cal{O}}^{ij}({\bf{p}})|\leq 1$, and so are also all of the matrix element of the $T_js$, for all $j=1,\dots,N$. One can therefore proceed to a $|z|\ll1$-expansion of (\ref{6}) that reads \cite{tgrh},
\begin{equation}\label{7}
\pm  (-{4\pi^2 m^2\over E^2})^N<\prod_{i=1}^N\, [1-i(-1)^{q_i}]\left(A_i+B_iz^{\frac{1+2q_i}{4}}+C_i{\sqrt{z}}\right)\left(1+(\sum_{h=1}^3O_{ih})\,z++{\cal{O}}(z^2)\right)>\,,
\end{equation}where $A_i,B_i, C_i,O_{ih}$ are pure $q_i$-dependent numbers. Upon averaging over $O_N({\mathbb{R}})$, the odd powers of $({\cal{OT}})_i$ vanish, but there remains contributions of order $z^0$, ${\sqrt{z}}$, $z$ and $z{\sqrt{z}}$ that are on the order of $({\cal{OT}})_i^0$, $({\cal{OT}})_i^2$, $({\cal{OT}})_i^4$ and $({\cal{OT}})_i^6$ respectively, and are leading in view of the smallness of $\lambda$, and/or $|z|$.
\par\medskip
Random orthogonal matrices can be generated in different ways, distributed according to the {\textit{Haar measure}} over the orthogonal group $O_N(\mathbb{R})$,~\cite{Anderson et al.}. An orthogonal matrix can be conveniently decomposed into a product of $N(N-1)/2$ {\textit{rotators}}, plus reflections, 
\begin{equation}\label{Eq:O}
{\cal{O}}=(R_{12}R_{13}\dots R_{1N})\,(R_{23}R_{24}\dots R_{2N})\dots \dots (R_{N-1,N})\,D_{{\mathbf{\varepsilon}}}\,,
\end{equation}
where the matrix of reflections is diagonal by definition and reads, $D_\varepsilon=diag(\varepsilon_1, \varepsilon_2,\dots,\varepsilon_N)$, with $\varepsilon_i=\pm 1,\ \forall i=1,\dots, N$. A random orthogonal matrix requires that to either value $\varepsilon_i=\pm 1$ be associated an equal probability $P(\varepsilon_i=\pm 1)=1/2$. A rotator $R_{ij}(\Theta_{ij})$, itself an $N\times N$-orthogonal matrix, acts as a rotation in the $(i\!-\!j)$-$2$-plane solely, and is thus characterized by an angle, $\Theta_{ij}$. The $\Theta_{ij}$ are independent random variables with a joint probability distribution proportional to~\cite{Anderson et al.},
\begin{equation}\label{angles}
\prod_{j=2}^N\cos^{j-2}\Theta_{1j}\,\prod_{j=3}^N\cos^{j-3}\Theta_{2j}\dots\prod_{j=N}^N\cos^{j-N}\Theta_{N-1,j}\,,
\end{equation}
whereas the probability density of an angle $\Theta_{ij}$ is a {\textit{beta}} distribution, $\beta(x_{ij};\frac{a}{2},\frac{b}{2})$, with $\cos\Theta_{ij}={\sqrt{x_{ij}}}$, that is, $\beta(x_{ij};\frac{a}{2},\frac{b}{2})=x_{ij}^{a/2\,-1}(1-x_{ij})^{b/2\,-1}/B(\frac{a}{2},\frac{b}{2})$. As meant in the second line of (\ref{Eq:1}), these probability densities allow one to calculate averages over orthogonal matrices in a definite quantitative way. However, for our present purpose, the full explicit form of the Haar measure is not required, but only the $D_\varepsilon$ matrix properties, so as the {\textit{left-}} and {\textit{right- invariances}} of the Haar measure on $O_N(\mathbb{R})$. Denoting with brackets those averages, and by $a^{ij}=a^{ij}(\dots,\Theta_{lm}\,,\dots)$ the matrix elements of (\ref{Eq:O}) as the reflection matrix $D_\varepsilon$ is omitted, one obtains,
\begin{equation}\label{2}
<{\cal{O}}^{ij}{\cal{T}}_j{\cal{O}}^{ik}{\cal{T}}_k>_{\varepsilon,\Theta}=<\varepsilon_j a^{ij}(\Theta)\,\varepsilon_ka^{ik}(\Theta){\cal{T}}_j{\cal{T}}_k>_{\varepsilon,\Theta}={\delta_{jk}}<a^{ij}(\Theta)a^{ik}(\Theta)>_\Theta{\cal{T}}_j{\cal{T}}_k\,,
\end{equation}where, in the last equality, the average over the product of reflections $\varepsilon_j\varepsilon_k$ has been taken. That is, $<{\sqrt{z}}>$ is given by
\begin{equation}\label{12}
\sqrt{\lambda}\sum_{j,k}^N<{\cal{O}}^{ij}{\cal{T}}_j{\cal{O}}^{ik}{\cal{T}}_k>_{\varepsilon,\Theta}= \sqrt{\lambda}<\sum_{j=1}^Na^2_{ij}(\dots\Theta_{lm}\dots)>_\Theta {\cal{T}}_j^2 \equiv \sqrt{\lambda}K_{\frac{1}{2}}\sum_j{\cal{T}}_j^2=\sqrt{\lambda}K_{\frac{1}{2}}\,DC_{2f}\, {\mathbf{1}}_{3\times 3}\,.
\end{equation}As a straight forward consequence  of the left- and right- invariances of the Haar measure on $O_N(\mathbb{R})$,~\cite{tgrh, {Anderson et al.}}, the second equality defines independently of the entry labels $i,j$ the constant $K_{{1}/{2}}$ as the mean value $<a^2_{ij}(\dots\Theta_{lm}\dots)>_\Theta$, {\textit{i.e.}}, $K_{{1}/{2}}=1/N$. 
Finally, $C_{2f}$ stands for the quadratic Casimir operator eigenvalue on the fundamental representation, $C_{2f}=C_F=4/3$.
In the same way, for the next sub-leading piece of (\ref{7}), one gets,
\begin{equation}\label{13}
<z>\,={\lambda}K_{1}\left(\,(DC_{2f})^2+(DC_{3f})\right){\mathbf{1}}_{3\times 3}\,,
\end{equation}where $K_{1}$ stands for the averages $<a^2_{ij}a^2_{ik}>_\Theta$, with, here, $K_{1}=1/N^2$ \cite{tgrh}. In (\ref{13}), one notices the second (cubic) Casimir operator eigenvalue over the fundamental representation of $SU_c(3)$,
\begin{equation}\label{cubic}
\sum_{a,b,c=1}^{N_c^2-1} d_{abc}\,t^at^bt^c\equiv C_{3f}{\mathbf{1}}_{3\times 3}\ .
\end{equation}The fully symmetric constants $d_{abc}$ are defined in the standard way, that is, in the case of interest, at $N_c=3$, $\{t_a,t_b\}=d_{abc}t_c+\frac{1}{3}\delta_{ab}$. Far less popular than $C_{2}$, the cubic Casimir operator eigenvalue over a representation space specified by the {\textit{Young Tableaux}} parameters $(p,q)$ is given by  
\begin{equation}\label{C3}C_{3}(p,q)=\frac{1}{18}(p-q)(2p+q+3)(2q+p+3),\end{equation}
and $C_{3f}=C_3(1,0)=10/9$ over the $SU(3)$ fundamental representation~\cite{Anirban}, whereas it is zero over the adjoint representation, $(1,1)$, another salient feature of the distinction between QCD and pure Yang-Mills.
At next sub-leading order, $z\sqrt{z}$, corresponding to the $O_N(\mathbb{R})$-averaged value of $({\cal{O}}^{ij}({\mathbf{p}}){\cal{T}}_j)^6$, calculations become more intricate; with $K_{{3}/{2}}=N^{-3}$,
\begin{eqnarray}\label{Eq:k11}
&& <z\sqrt{z}>\,= \lambda^{\frac{3}{2}}K_{\frac{3}{2}}\,\biggl\lbrace\left(\,2(DC_{2f})^2+(DC_{2f})(DC_{3f})+\frac{4}{3}(DC_{3f})\right){\mathbf{1}}_{3\times 3}\nonumber\\ && +\sum_{k,j,l,h,m}\, d_{kjm}d_{khl}\, (T_jT_mT_hT_l+2 T_jT_hT_lT_m)\biggr\rbrace\,.
\end{eqnarray}The two last terms, somewhat puzzling, seem to compromise the general structure of these dependences. To proceed, one may rely on the standard values of the $d_{abc}$ coefficients~\cite{Yndurain}. Then, working out identities such as,
\begin{equation}\label{identity}\sum_{k,j=1}^8\, d_{kjj}=0,\ \ \ \ \ \ \ 
\sum_{j,m=1}^8\, d_{k'jm}d_{kjm}=\frac{5}{3}\,\delta_{k'k}\,,
\end{equation} one can prove that (\ref{Eq:k11}) indeed reduces to,
\begin{equation}\label{18}
<z\sqrt{z}>\,= \lambda^{\frac{3}{2}} K_{\frac{3}{2}}\left(\,[2+(\frac{5}{6})^2](DC_{2f})^2+(DC_{2f})(DC_{3f})+{3}(DC_{3f})\right){\mathbf{1}}_{3\times 3}\,.
\end{equation}\

\section{\label{SEC:3} Conclusion}

 With Eqs.(\ref{12}), (\ref{13}) and (\ref{18}), the point of the present letter is reached. At quenched and eikonal approximations at least, the strong coupling QCD fermionic amplitudes exhibit a behaviour that, contrarily to usual results, is no longer dependent on the first, quadratic Casimir operator alone, $C_{2f}{\mathbf{1}}_{3\times 3}$, but also on the second, cubic Casimir operator, $C_{3f}{\mathbf{1}}_{3\times 3}$. This feature, given the rank-2 character of the $SU_c(3)$-color algebra, had to come into play in one way or another. It is here made visible on the generic form of the strong coupling fermionic QCD amplitudes such as derived in Ref.\cite{QCD6}, thanks to the property of effective locality~\cite{QCD1,QCD-II}. That property, which goes along with a mass scale~\cite{QCD6}, is non-perturbative, gluonic degrees of freedom being integrated out. Note that on the same numerical bases as discussed after (\ref{D4}), and at $\lambda$ close to 1 (a heavy quark system, {\textit{e.g.}}), (\ref{13}) can reach some $20\%$ of (\ref{12}); and there, in (\ref{13}), the $C_{3f}$ contribution, about $16\%$ of ${C_{2f}}'s$ one, while it goes up to $48\%$ in (\ref{18}).
 \par
  It is interesting to note that effective locality leads to a description of any $2n$-point fermionic Green's function in terms of $C_{2f}$ and $C_{3f}$, which, in  {\textit{Quark Models}} without gluons, correspond to $2$- and $3$-body potential interactions, with the latter bringing about significant improvements to a description based on a sole $2$-body potential~\cite{C3}. As is well-known in atomic and nuclear physics, that point seems to be generic of a variety of similar situations: In the case of {\textit{shallow systems}} it has been shown that, at leading order, the only potentials that are necessary to a complete description of an $n$-body system, are the $2$- and $3$- body interaction potentials; and that the sole $2$-body potential is not enough, whatever the parametrizations one tries to complete that $2$-body description with~\cite{Mario}.
 \par\medskip
 Further on, it will be worth analyzing how the eikonal and quenching approximations affect those Casimir operator dependences. Though consistent with a non-perturbative regime, it will be interesting also to relax the strong coupling limit which here, as in Ref.\cite{QCD6}, is introduced in a somewhat `academic way', so as to get rid of a subleading pure Yang Mills term, which so far at least, couldn't be treated on the same footing as all of the other terms~\cite{{QCD6},tgrh}; that is, by means of a random matrix calculation.

\begin{acknowledgments}
It is a pleasure to thank H.M. Fried for a careful reading of the manuscript, M. Gattobigio for having provided references most relevant to the subject, and R. Hofmann who pointed out two errors.
\end{acknowledgments}

\par

\end{document}